\documentclass[12pt,preprint]{aastex}
\def\gta{ \lower .75ex \hbox{$\sim$} \llap{\raise .27ex \hbox{$>$}} }
\def\lta{ \lower .75ex\hbox{$\sim$} \llap{\raise .27ex \hbox{$<$}} }
\def \Ol {$\Omega_\Lambda$}
\def \Om {$\Omega_{\rm M}$}

\begin{document}

\title{Gamma Ray Bursts: new rulers to measure the Universe}

\author{Giancarlo Ghirlanda\altaffilmark{1}, 
Gabriele Ghisellini\altaffilmark{1},
Davide Lazzati\altaffilmark{2} and 
Claudio Firmani\altaffilmark{1}}
\affil{1 INAF -- Osservatorio  Astronomico di  Brera, via  Bianchi 46,
23807 Merate, Italy} 
\affil{2 Institute of  Astronomy, Madingley Road CB3 0HA, Cambrige UK}

\setcounter{footnote}{0}

\begin{abstract}
The best measure of the Universe should be done using a standard
``ruler" at any redshift.  Type Ia Supernovae (SN Ia)
probe the universe up to z$\sim$1.5, while the 
Cosmic Microwave Background (CMB) 
% generated at $z\sim$1000, carries information on the intermediate
% redshift Universe only through integrated effects between 
% $z\sim 1000$ and the observer.
primary anisotropies concern basically
$z\sim$1000.
Apparently, Gamma--Ray Bursts (GRBs) are all but standard candles.  
However, their emission is
collimated and the collimation--corrected energy correlates tightly
with the frequency at which most of the radiation of the prompt is
emitted, as found by Ghirlanda et al. (2004).  
Through this correlation we can infer the burst 
energy accurately enough 
to probe the intermediate redshift ($z<10$) Universe.
Using the best known 15 GRBs we find very encouraging results
that emphasize the cosmological GRB role. 
A combined fit with SN Ia
yields $\Omega_{\rm M}=0.37\pm0.10$ and $\Omega_{\Lambda}=0.87\pm 0.23$.
Assuming in addition a flat Universe, the parameters are 
constrained to be $\Omega_{\rm M}=0.29\pm0.04$ and 
$\Omega_{\Lambda}=0.71\pm 0.05$.
GRBs accomplish the role of ``missing link" between SN Ia and CMB primary
anisotropies. 
They can provide a new insight on the cosmic effects of dark
energy, complementary to the one supplied by CMB secondary anisotropies through
the Integrated Sachs Wolfe effect. 
The unexpected Standard Candle cosmological
role of GRBs motivates the most optimistic hopes for what can be obtained
when the GRB-dedicated satellite, Swift, will be launched.

%
% Furthermore, probing the Universe with high accuracy up to
% high redshifts, GRBs establish a new insight on the cosmic expanding
% acceleration history.
% GRB accomplish the role of ``missing link" between the CMB
% and SN Ia, motivating the most optimistic hopes for what can 
% be obtained when the GRB-dedicated satellite, Swift, will be 
% launched.
\end{abstract}


\section{Introduction}

Recently, Ghirlanda, Ghisellini and Lazzati (2004, GGL04 thereafter)
found a surprisingly tight correlation between the peak of the
$\gamma$--ray spectrum $E_{\rm peak}$ (in a $\nu F_\nu$ plot) and the
collimation corrected energy emitted in $\gamma$--rays $E_\gamma$ 
in long Gamma-Ray Bursts (GRBs).  The latter is related to the
isotropically equivalent energy $E_{\gamma, iso}$ by the value of the
jet aperture angle $\theta$, by $E_\gamma = E_{\gamma, iso}
(1-\cos\theta)$.  The scatter around this correlation is tight
enough to wonder if the correlation itself can be used for a reliable
estimate of $E_{\gamma, iso}$, making GRBs distance indicators,
and therefore probes for the determination of the cosmological \Om,
\Ol ~parameters, and for the exploration of the matter to vacuum 
dominance transition.

This issue is similar to the case of SN Ia: they are not perfect standard
candles (i.e. their luminosities are not all the same), nevertheless
the luminosity of a specific supernova can be found through the
correlation of their luminosity and the speed of the decay of their
light curve (i.e. the slower the brighter, Phillips 1993; Riess, Press \&
Kirshner 1995).  It is the existence of this correlation among SN Ia
which made possible their cosmological use (Riess et al. 2004,
hereafter R04; Perlmutter et al. 1999; Schmidt et al. 1998).

Very recently, this problem has been explored by Dai, Liang \& Xu
(2004), which found tight constraints on \Om ~and \Ol ~using the
correlation found by GGL04. 
Their result is however based on a strong assumption: 
they assume as universal the correlation 
measured in a particular cosmology  
(without errors on its slope),
and use it to derive the cosmology itself. 
Actually, the best fit correlation depends on the cosmology 
adopted to derive burst luminosities and the correlation
should be re--calibrated for each cosmology. 

In this letter we demonstrate that a correct approach leads to
less tight constraints on the cosmological parameters using GRBs
alone. On the other hand, a more interesting cosmological 
result can be acheived if a combined fit with SN Ia is performed.

\section{The Hubble diagram of GRBs}

As in the case of SN Ia, the use of a class of objects as cosmological
``rulers'' requires  them being  standard candles.  The  luminosity of
GRBs, calculated assuming isotropy, spans $\sim 4$ orders of magnitude
(Frail  et al.  2001),  but strong  observational  evidence (i.e.  the
achromatic  break in  the afterglow  light--curve) indicates  that the
burst emission  is collimated into  a cone/jet of some  aperture angle
$\theta$  (Levinson \& Eichler 1993; Rhoads  1997; Sari,  Piran  \&  
Halpern 1999; Fruchter  et al.  1999).   
The  corresponding  energy  emitted  in  $\gamma$--rays,
corrected by the  collimation factor $(1-\cos\theta)$, clusters around
$E_\gamma\sim 10^{51}$ erg, with a small dispersion (0.5 dex), yet not
small enough for a cosmological use (Bloom et al. 2003).

GGL04 found a tight correlation between $E_\gamma$
and the (rest frame) peak energy $E_{\rm peak}$ of the $\nu F_\nu$ prompt
emission spectrum: $E_\gamma \propto E_{\rm peak}^x$.  The exact value of
$x$ depends on the assumed cosmology.  
Using \Om$=0.3$ and \Ol$=0.7$ we have
\begin{equation}
E_\gamma \, =\, (4.3\pm 0.9)\times 10^{50}
\left({E_{\rm peak}\over 267 ~~{\rm keV}}\right)^{1.416\pm 0.09}
\quad {\rm erg}
\label{corr}
\end{equation}
The scatter of the data points around the correlation is of the order
of 0.1 dex.  This allows to reconstruct the value of $E_\gamma$ by
measuring $E_{\rm peak}$.

This is analogous to SN Ia, for which there is a tight relation
between their peak luminosity and the stretching factor of their
optical light curve (Phillips 1993; Goldhaber et al. 2001), with less
luminous supernovae showing a faster post--maximum light curve decay
(Reiss et al. 1995).  The proper modelling of this effect (Hamuy et
al. 1996; Perlmutter et al. 1999) improves the determination of the
SN Ia luminosity and consequently reduces the scatter in the Hubble
diagram, yielding constraints on the cosmological parameters (see R04
using SN Ia with redshift up to $z\sim$1.75).

The $E_\gamma - E_{\rm peak}$ correlation for GRBs makes them a
new class of ``rulers'' for observational cosmology and combining GRBs
and SN Ia can further reduce the region of allowed values in the
cosmological parameter space.  Furthermore, since GRBs are detectable
at larger $z$, they are a powerful tool to explore in more detail the
cosmic kinematics.

The difference between the standard candle assumption and the use of
the intrinsic correlations, for both GRBs and SN Ia, is shown in Fig. 1
(top and bottom panel, respectively) through the Hubble diagram in the
form of luminosity--distance vs redshift.  In the upper panel we
assume that GRBs and SN Ia are standard candles with a unique energy
($E_\gamma=10^{51}$ erg) for GRBs and luminosity ($B$=-21.1) for SN.
The derivation of the luminosity distance $D_{\rm L}$ for SN follows 
straightforwardly from their distance modulus (R04).  
For GRBs we have 
$D_{\rm L} \equiv (1+z)E_\gamma/[4\pi {\cal F_\gamma} (1-\cos\theta)]$, where
${\cal F_\gamma}$ is the $\gamma$--ray fluence (i.e. the time
integrated $\gamma$--ray flux).  
Note that the determination of
$\theta$ requires the knowledge of the isotropic energy (see e.g.
Eq. 1 in Frail et al. 2001), 
in turn requiring specific values of (\Om, \Ol).  
In the bottom panel we plot SN Ia and GRBs after correcting for the 
stretching--luminosity and the $E_{\gamma}$--$E_{\rm peak}$ relations, 
respectively.  
In this case the isotropic energy $E_{\gamma, iso}$ of GRBs has 
been estimated from their measured $E_{\rm peak}$ through the 
$E_\gamma$--$E_{\rm peak}$ correlation and the error on the slope 
of this correlation has been properly included in the $D_{\rm L}$ 
total uncertainty.
Also in this case we must fix a given (\Om, \Ol)
cosmology both for the derivation of $\theta$ and for finding the best
$E_{\gamma}$--$E_{\rm peak}$ relation.  
As in the SN Ia case, the
luminosity distance of GRBs derived from their $E_\gamma$--$E_{\rm peak}$
correlation (bottom panel) highly reduces the scatter around possible
different cosmologies (solid, dashed and dotted lines).  
Moreover, GRBs populate the $z>1$ region, where $D_{\rm L}(z)$ is rather 
sensitive on (\Om, \Ol).

\section{Constraints on cosmological parameters}

The correlation found by GGL04 {\it assumes} \Om=0.3 and \Ol=0.7, and
$h=0.7$.  Changes on \Om ~and \Ol ~induce a change on the normalization
and slope of Eq. \ref{corr}, together with a {\it different scatter}
of the data points around the best fit line.  We can then ask what is
the pair of cosmological values \Om, \Ol ~which produces the ``minimum
scatter'' around the fit, performed using the very same \Om, \Ol ~pair.
To this aim we use all the 15 bursts of known redshifts, $E_{\rm peak}$ 
and jet break time $t_{\rm break}$ listed in Tab. 1 and Tab. 2
of GGL04.

The difference with the study of Dai, Liang \& Xu (2004) lies mainly
in this point: they assumed that the $E_\gamma$-$E_{\rm peak}$
correlation is exact and cosmology independent, while it is not
\footnote{ 
For instance, using \Om=0.4, \Ol=0.6 ~results in
$E_\gamma=(3.7\pm 0.9)\times 10^{50} (E_{\rm peak}/267 ~~{\rm keV})^{1.38\pm 0.09}$ erg
(i.e. a $\sim$2.6\% and $\sim$16\% change in slope and normalization
with respect to Eq. 1).
% Using \Om=0.6, \Ol=0.4 we have 
% $E_\gamma=(3.4\pm 0.9)\times 10^{50} (E_{\rm peak}/267 ~~{\rm keV})^{1.34\pm 0.09}$ erg
% (i.e. a $\sim$5.7\% and $\sim$26\% change in slope and normalization
% with respect to Eq. 1.
With \Om=1, \Ol=1 we obtain
$E_\gamma=(3.0\pm 0.9)\times 10^{50} (E_{\rm peak}/267 ~~{\rm keV})^{1.29\pm 0.08}$ erg
(i.e. a $\sim$9\% and $\sim$30\% change in slope and normalization
with respect to Eq. 1).
}.

Additional differences concern: 
i) the estimate of the errors on the
density of the interstellar medium when it is unknown (they assume
$n=3\pm 0.33$ cm$^{-3}$ while we allow $n$ to cover the entire 1--10
cm$^{-3}$ range); 
ii) we do not exclude GRB 990510 and GRB 030226 from
the analysis; 
iii) we include GRB 030429, for which a jet break time
was recently found by Jakobsson et al. (2004); 
iv) we always use $(1-\cos\theta)$ (instead of the $\theta^2/2$ 
approximation) as the collimation correction factor (also when
estimating the error on $E_\gamma$).

We also consider the 156 SN Ia of the ``Gold'' sample of R04, finding the
corresponding \Om, \Ol ~contours using the distance moduli and
corresponding errors listed in their Tab. 5.  
Fig. 2 shows our results.  
GRBs alone (red lines) 
are almost insensitive to \Ol ~but limit \Om ~to lie within
$\sim$0.05 and 0.22 (68\% confidence level).

We also show the region pinpointed by the WMAP experiment 
(Spergel et al. 2003), which
is only marginally consistent with the allowed region from SN Ia alone
(blue lines in Fig. 2).  The combined GRBs+SN Ia fit (filled regions in
Fig. 2) selects a region which is more consistent with the cosmic
microwave background (CMB) results.  
The minimum (with a reduced $\chi_{red}^2$=1.146) is for 
\Om=0.37$\pm0.15$ and \Ol=0.87$\pm0.23$ (1--$\sigma$).  
Assuming a flat universe yields \Om=0.29$\pm0.04$ and
\Ol=0.71$\pm0.05$.

If we use the ``classical" Hubble diagram method, we compare the $D_{\rm L}$
values given by estimating $E_\gamma$ through the actual correlation
found in each point of the \Om, \Ol ~plane with the luminosity distance
calculated through e.g. Eq. 11 of R04 (see also Carrol et al. 1992).
Then, by a $\chi^2$ statistics, we find the confidence regions in the
\Om, \Ol ~plane, which are plotted as dashed line on Fig. 2.  This
classical method is very similar to the previous one, since it uses
the same available information.  The small difference (contours
slightly larger) is due to the fact that with the ``minimum scatter"
method we calculate the distance of the data points from the
correlation (i.e. perpendicular to the fitting line), while, using the
``classical" Hubble diagram method, we are using the distance between 
the $E_\gamma$ data point and the corresponding $E_\gamma$ by the 
correlation.

We can further constrain, with the combined GRB and SN samples, the
dark energy component which is parametrized by its equation of state
$P=w\rho c^2$.  
Furthermore, $w$ could be varying, and one possible
parametrization is $w=w_{0}+w^\prime z$ (see e.g. R04). 
Adopting this law, we
compute the luminosity distance according to Eq. 14 of R04,
assuming a flat cosmology with \Om=0.27.
In this case the fit is performed in the $w_{0}$--$w^\prime$ plane 
for GRBs, SN and for the combined samples.  
As before, we recompute the  $E_\gamma$--$E_{\rm peak}$
relation for each $w_0$, $w^\prime$ pair
\footnote{
As an example of how the correlation is sensitive to the change
of $w_0$, $w^\prime$, consider that, for $w_0=-0.7$ and $w^\prime=0.2$
the correlation becomes
$E_\gamma=(3.75\pm 0.90)\times 10^{50} (E_{\rm peak}/267 ~~{\rm keV})^{1.37\pm 0.09}$ erg
(i.e. a $\sim$3.4\% and $\sim$15\% change in slope and normalization
with respect to Eq. 1).
}. 
Fig. 3 reports the corresponding confidence intervals.
Besides making the confidence region smaller than what derived for SN
alone, the effect of GRBs is to include within the 68\% contour 
of the joint SN+GRB sample (filled region) the $w_0=-1$, $w^\prime=0$ 
point, corresponding to the classical cosmological constant.

\section{Discussion}

GRBs can now be used as cosmological ``rulers", 
bridging the gap between the relatively nearby type Ia
supernovae and the cosmic microwave background.  The SWIFT satellite
(Gehrels et al. 2004), designed for the fast localization of GRBs, is
expected to find about one hundred of GRBs per year: it can open up a
new era for the accurate measurements of the geometry and kinematics
of our Universe (for a more extended discussion see Firmani et al. 2004).  
We stress that, besides finding high redshift
bursts, which are of course very important for finding tighter
constraints, it is crucial to find {\it low redshift} GRBs, to determine
the $E_\gamma$--$E_{\rm peak}$ correlation in a redshift range which is
not affected by the \Om, \Ol ~values.  This would allow to use the
resulting correlation unchanged for all values of \Om, \Ol, strongly
reducing the associated errors. 
In turns, this will allow to constrain cosmological 
parameters independently from SN Ia. 
This is important since GRBs are unaffected by dust extinction and
it is very unlikely that two completely different classes of objects
would have similar evolutions to mimic a consistent set of
cosmological parameters.

In Fig. 4 we show an illustrative example of what can be done {\it if
a given correlation were known to be valid independently of the
cosmological parameters}.  For this we have chosen the correlation
given by Eq. \ref{corr}.  It can be seen that even the limited sample
of our bursts can strongly influence the GRB+SN confidence contours,
making them more in agreement with the WMAP results (not unexpectedly,
since we have used just the correlation appropriate for \Om=0.3 and
\Ol=0.7. A similar consideration concerns the Dai et al. 2004 result).

We would like to stress that in order to use GRBs to find the
cosmological parameters, we need a set of well measured data, and
especially a well measured jet break time $t_{\rm break}$, necessary to
find the collimation angle $\theta$, and a good spectral determination
of the prompt emission.

\acknowledgments{We thank Annalisa Celotti for useful discussions.  DL
thanks the Osservatorio Astronomico  di Brera for the kind hospitality
during part of the preparation  of this work.  G. Ghirlanda thanks the
MIUR for COFIN grant 2003020775\_002.}

\clearpage
\newpage
\clearpage

\begin{figure}
\begin{center}
\resizebox{!}{7in}
% {\vsize}
%\hskip{1.5cm}
{\includegraphics{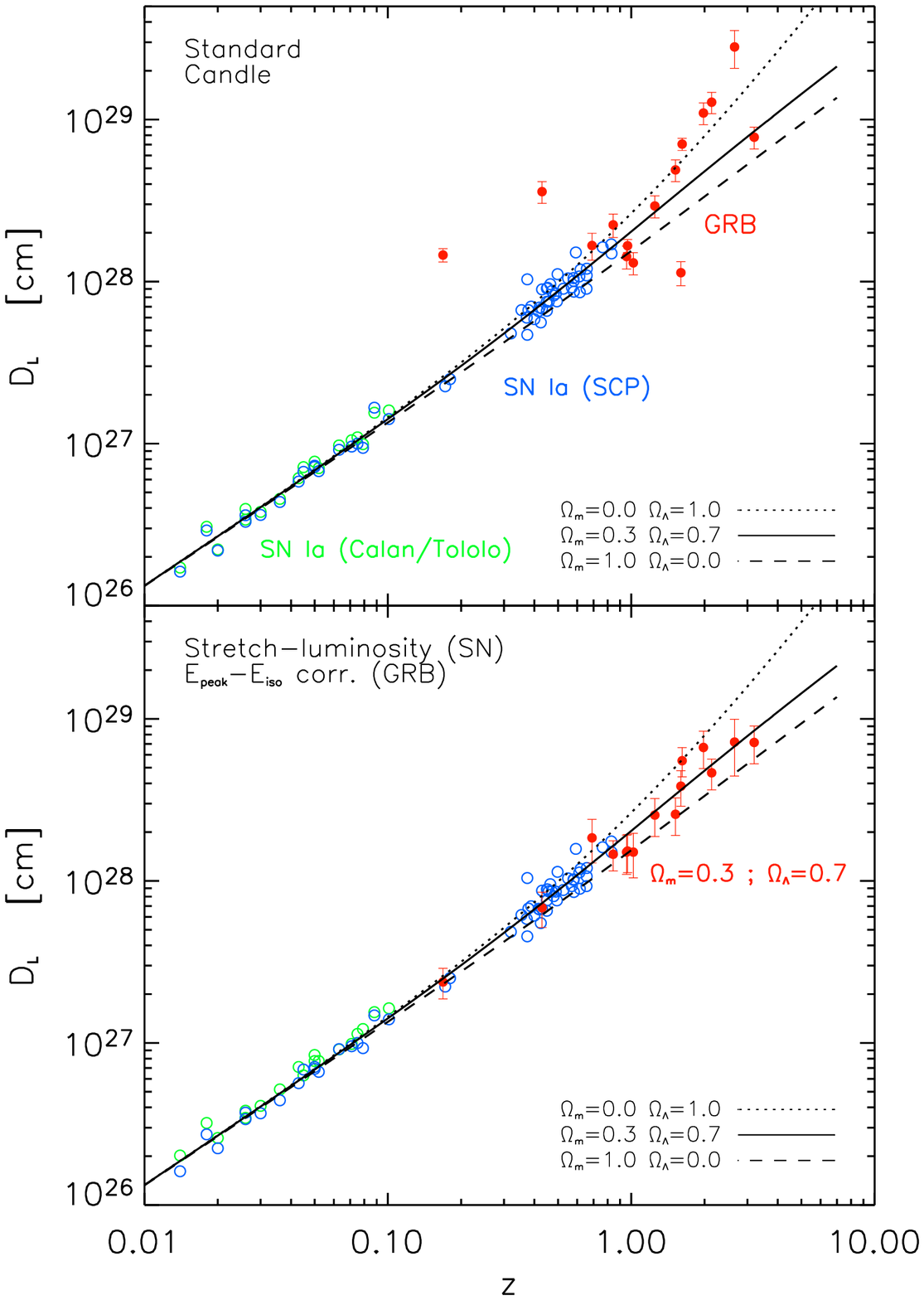}}
\figcaption[]{
Classical Hubble  diagram in the
form of  luminosity--distance $D_{\rm L}$ vs redshift $z$ for Supernova Ia
(open  green circles: Cal\`an/Tololo  sample (Hamuy et al. 1996);  
open blu circles:  Perlmutter  et  al. 1999) 
and GRBs (filled  red circles: the 15 bursts in GGL04).  
In the top panel the
SN Ia  and  GRBs  are treated as  standard  candles  (no  corrections
applied); for GRBs $E_\gamma=10^{51}$ erg is assumed.  In the
bottom  panel  we  have  applied the  stretching--luminosity  and  the
$E_\gamma$--$E_{\rm peak}$ relations  to SN Ia and  GRBs, respectively, as
explained in  the text.  Note  that, for GRBs, the  applied correction
depends upon  the specific assumed cosmology: here for  simplicity we
assume the standard \Om=0.3, \Ol=0.7 ~cosmology.  Both panels also show
different $D_{\rm L}(z)$ curves, as labelled.
\label{fig1}}
\end{center}
\end{figure}

\clearpage

%-----------------------------------------------
\newpage
\clearpage

\begin{figure}
\begin{center}
\resizebox{17cm}{17cm}
% {\vsize}
{\includegraphics{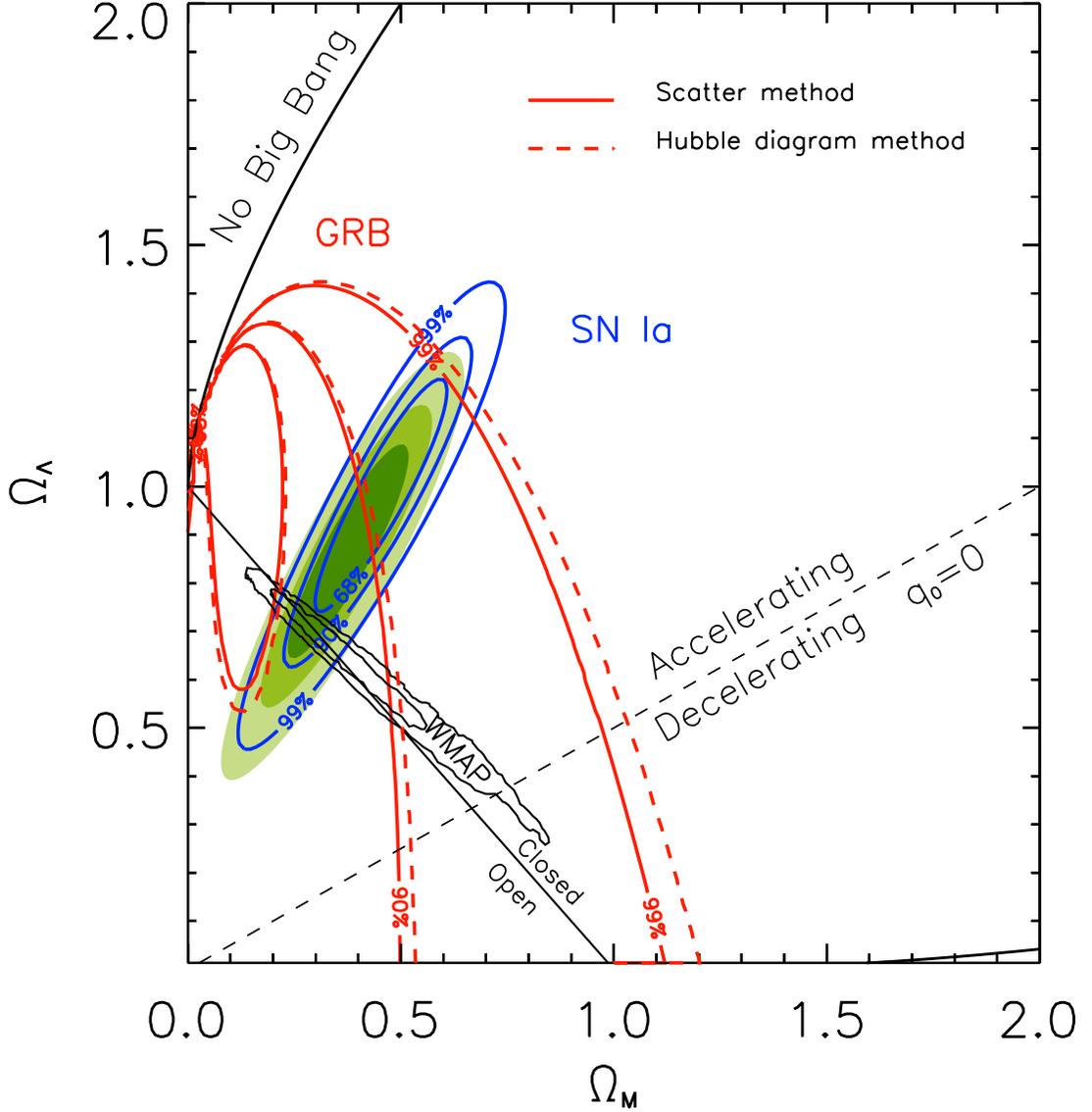}}
\figcaption[]{
Constraints in the \Om--\Ol  ~plane derived for 
our GRB sample (15 objects, red contours); 
the ``Gold" Supernova Ia sample of R04
(156 objects, blue contours, derived assuming a 
fixed value of $H_0=65$ km s$^{-1}$ Mpc$^{-1}$, 
making the contours slightly different from Fig. 8 of R04).
The WMAP satellite constraints (black contours,
Spergel et al. 2003) are also shown.
The three colored ellipsoids are the confidence
regions (dark green: 68\%; green: 90\%; light green: 99\%) 
for the combined fit of SN Ia and our GRB sample.
For GRBs only, the miminum $\chi_{red}^2=1.04$, is at \Om=0.07, \Ol=1.2.
\label{fig2}}
\end{center}
\end{figure}

\clearpage
%-----------------------------------------------
\newpage
\clearpage

\begin{figure}
\begin{center}
\resizebox{17cm}{17cm}
% {\vsize=10 true cm}
{\includegraphics{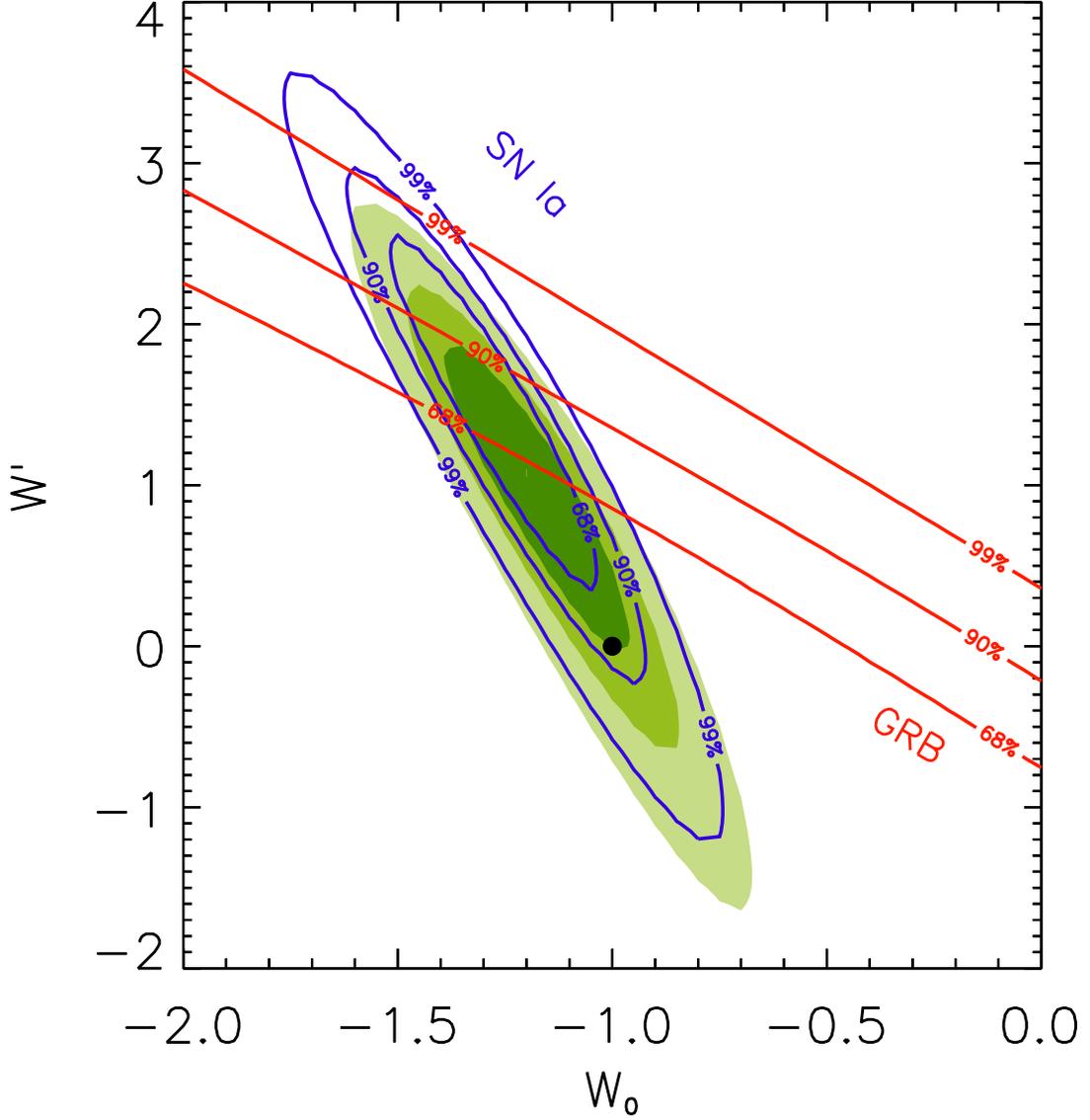}}
\figcaption[]{Constraints on the $w_0$, $w^\prime$ parameters
entering the equation of state $p=(w_0+w^\prime z)\rho c^2$,
where $\rho$ is the dark energy density.
$w_0=-1$ and $w^\prime=0$ correspond to the
cosmological constant \Ol.
We assume a flat geometry and \Om=0.27 (see also R04).
Blue contours: constraints from type Ia SN (R04).
Red contours: constraints from our GRBs,
Colored regions: combined constraints (dark green, green and light green
for the 68\%, 90\% and 99\% confidence levels, respectively).
\label{fig3}}
\end{center}
\end{figure}

\clearpage
%-----------------------------------------------
\newpage
\clearpage

\begin{figure}
\begin{center}
 \resizebox{16.5cm}{16.5cm}
% {\vsize}
{\includegraphics{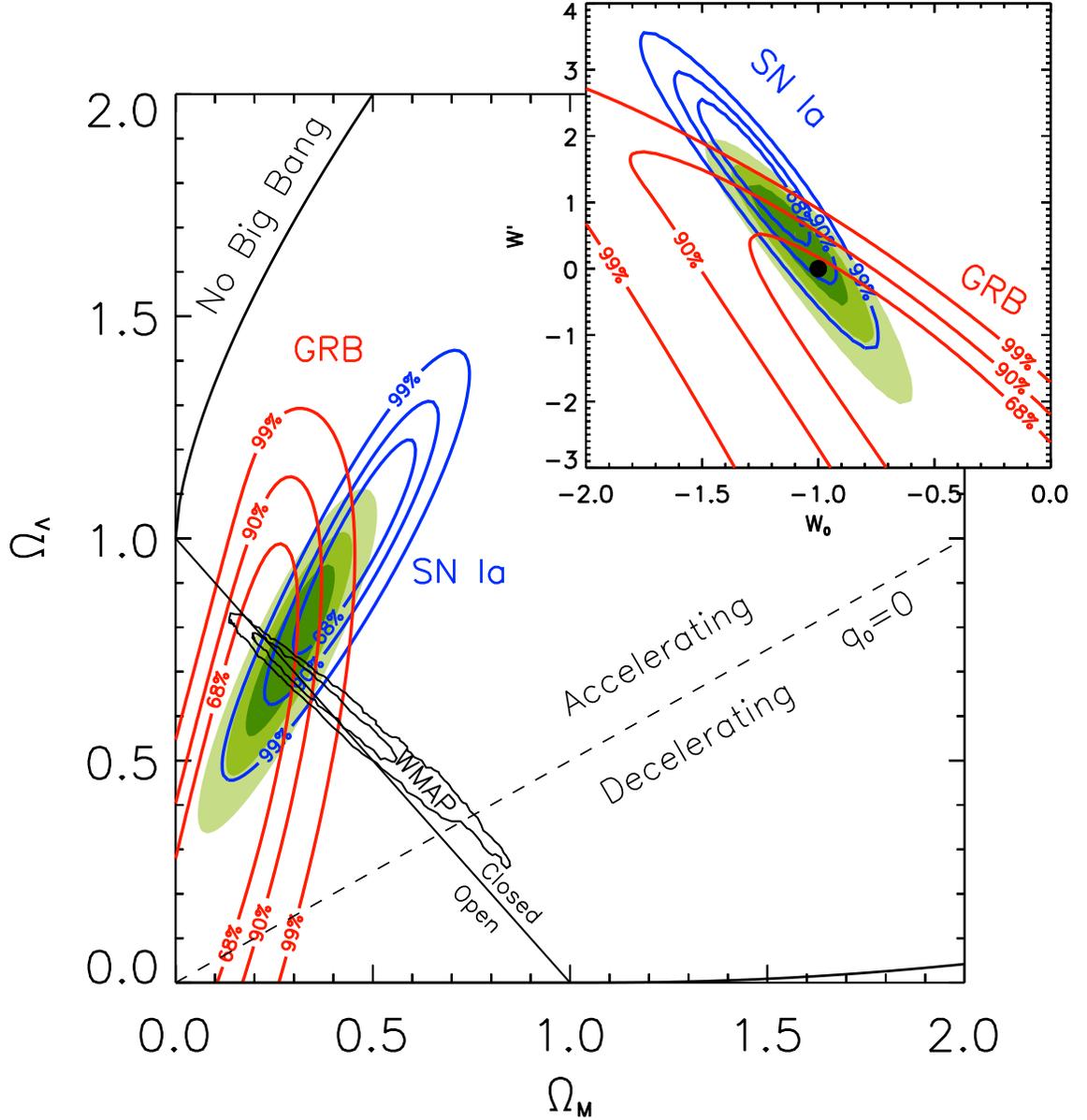}}
\figcaption[]{
Example of how GRBs can contribute to the determination of the cosmological
parameters once the $E_\gamma - E_{\rm peak}$ correlation will be
found in a cosmology independent way (i.e. finding bursts
at small redshifts).
For this example we assume that the correlation of Eq. \ref{corr} 
is valid for any cosmological parameter.
We show the contours in both the \Om, \Ol ~plane
(main figure) and in the $w_0-w^\prime$ plane (insert,
a flat cosmology with \Om=0.27 is assumed).
Lines and colors are as in Fig. 2 and Fig. 3. 
\label{fig4}}
\end{center}
\end{figure}

\end{document}